\documentclass[pre,aps,pacs,twocolumn]{revtex4}
\usepackage{amsmath,bm,epsfig}
\usepackage{color}

\usepackage[latin1]{inputenc}
\newcommand{\beq}{\begin{equation}}
\newcommand{\eeq}{\end{equation}}
\newcommand{\bea}{\begin{eqnarray}}
\newcommand{\eea}{\end{eqnarray}}
\newcommand{\ud}{\mathrm{d}}
\begin{document}
\title{Statistical Mechanics of Glass Formation in Molecular Liquids with OTP as an Example}
\author{Laurent Bou\'e, H.G.E. Hentschel, Valery Ilyin and  Itamar Procaccia}
\affiliation{The Department of Chemical Physics, The Weizmann
Institute of Science, Rehovot 76100, Israel}
\begin{abstract}
We extend our statistical mechanical theory of the glass transition from examples consisting of point particles
to molecular liquids with internal degrees of freedom. As before, the fundamental assertion is that super-cooled
liquids are ergodic, although becoming very viscous at lower temperatures, and are therefore describable in principle
by statistical mechanics. The theory is based on analyzing the local neighborhoods of each molecule, and a statistical
mechanical weight is assigned to every possible local organization. This results in an approximate theory that is in
very good agreement with simulations regarding both thermodynamical and dynamical properties.
\end{abstract}
\pacs{PACS number(s): }
\maketitle

\section{Introduction}

The glass transition is a rather dramatic phenomenon in which the viscosity of super-cooled liquids shoots up
some 14 or 15 orders of magnitude over a small range of temperature \cite{09Cav}. In fact, the phenomenon is so
extreme that many authors fitted their measured relaxation time to the Vogel-Fulcher formula which predicts a total
arrest of any relaxation at some finite temperature. Recent thinking does not support the existence of a finite temperature
singularity. In \cite{08EP} it was argued that in simple glass-forming models the system remains ergodic at all temperatures,
for any finite number of particles.  In \cite{11HKLP} it was shown that even at $T=0$ amorphous solids are not purely elastic,
having some viscous plastic response still there even in the thermodynamic limit. In a series of papers it was shown that
the subtle structural changes that occur in simple models of super-cooled liquids are quite well described by an approximate
statistical mechanics that is based on a reasonable up-scaling method that defines discrete partition sums involving
quasi-species of particles and their neighbors \cite{BLPZ09}. All these
indicate that it may be fruitful to proceed under the assumption
that the glass transition does not involve the breaking of ergodicity, and that therefore statistical mechanics can be
applied, in fact making use of the extremely slow relaxation to provide a quasi-equilibrium theory in spite of the
very slow aging \cite{Dhar}. The aim of this paper is to extend this approach to the glass transition in molecular
liquids. These provide additional challenges that do not exist in simple models involving point particles, mainly
due to the existence of  of supplementary degrees of freedom and of more
complex interactions between the basic
constituents. We will show, using the example of OTP (ortho-terphenyl) which is a very well known glass former whose
glass transition is extensively studied, that the program discussed above continues to make sense, providing very needed
insights into the phenomenology of molecular glass formers.

The structure of the paper is as follows: In Sect. \ref{system} we introduce the OTP system, explain the model used,
and present the simulation results that need to be understood. In Sect. \ref{statmech} the statistical mechanical theory
is introduced and explained, together with extensive comparisons of the predictions of the theory to the results of
simulations. In Sect. \ref{viscosity} we address in particular the temperature dependence of the viscosity of super-cooled OTP in
light of the proposed theory. Finally, Sect. \ref{discussion} is dedicated to a summary and a discussion of
the main results of this paper.

\section{OTP and its thermodynamics}
\label{system}
\subsection{The system}
One of the best studied glass-forming system is ortho-terphenyl
(1,2-diphenylbenzene), denoted for brevity OTP. In general, terphenyls (molecular formula
$C_{18}H_{14}$ with  molecular mass
$M=230.31 g/mol$) are a group of closely-related aromatic hydrocarbons. They
consist of a central benzene ring substituted with two phenyl groups. There
are three isomers,  ortho-, meta-, and para-terphenyl.

\begin{figure}[!h]
\centering
\epsfig{width=.38\textwidth,file=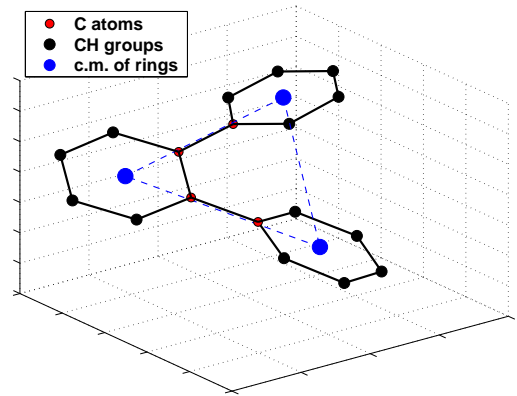}
\caption{Color online: The structure of the ortho-terphenyl molecule (atomic coordinates
from \cite{AMOS78}).}
\label{fig1}
\end{figure}

The structure of the OTP molecule was studied in Refs.
\cite{CL37,KB44,AMOS78,BL79} by different experimental methods  and is shown
in Fig. \ref{fig1}.
The melting temperature of OTP is $T_{m}=330 K$ and the boiling temperature is
$T_b=605 K$ \cite{BC27}. The  ``glass transition temperature" $T_{g}$ defined
by volumetric and thermal measurements (cf. Figs. \ref{fig3},\ref{fig5} below) is near $243 K$ \cite{GT67}
(This temperature depends on the thermal protocol by which
the glass is prepared \cite{CB72}).

\begin{figure}[!h]
\centering
\epsfig{width=.38\textwidth,file=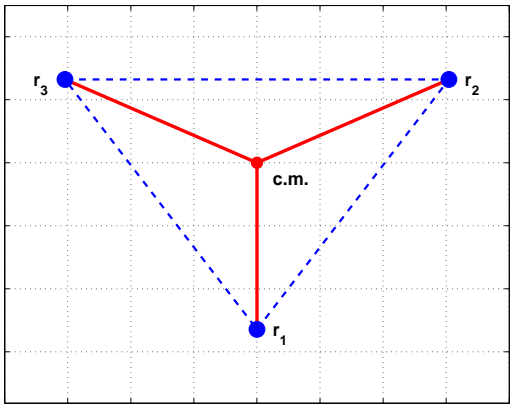}
\caption{Color online: The planar LW-model of the ortho-terphenyl molecule.}
\label{fig2}
\end{figure}

\subsection{The model employed in simulations}
A simple coarse grained rigid model of OTP, known as the Lewis-Wahnstr\"om
(L-W) model, was introduced for the first time in  Refs. \cite{LW93,LW94}. Some intensive
studies of the properties of this model were achieved by molecular dynamics (MD) simulations
\cite{LW93,LW94,RST01,MNSDST02,CS04,LDS06}. More realistic  representation
of the OTP molecule consists in taking into account internal degrees of
freedom and a number of models of such kind were developed and used in MD
simulations \cite{KW95,MLRS00,MLRS01,GF06,BRDS06}.

For the sake of efficient numerical simulations the full resolution of the structure of the OTP molecule
using a sufficiently large number of molecules is beyond our computer
capabilities. One needs to invoke a reasonable model that can retain the
essential aspects of the molecular structure while allowing reasonable
computations. Therefore we choose the simple Lewis-Wahnstr\"om  model and use it
for Monte Carlo (MC) simulations. The OTP molecule is represented by a
three-site complex, each site playing the role of the whole benzene ring
(see Fig. \ref{fig1}). The interaction between two sites of different
molecules is represented by a Lennard-Jones potential
\begin{equation}
\phi_{LJ}(r)=4\epsilon \Bigg[ \bigg(\frac{\sigma}{r}\bigg)^{12}-
\bigg(\frac{\sigma}{r}\bigg)^6\Bigg],
\label{lennard}
\end{equation}
where $r$ is the distance between sites. For the sake of simulation speed we employ a smooth short range version of the
interaction potential as proposed in \cite{MNSDST02}
\begin{equation}
\phi(r)=\left\{
\begin{array}{ll}
\phi_{LJ}(r)+a+b r&\textrm{if $r\le r_{cut}$} \\
0&\textrm{if $r > r_{cut}$}
\end{array} \right.
\label{ljshort}
\end{equation}
Here, $\epsilon=5.276$ kJ/mol and  $\sigma=0.483$ nm. The parameters $a$ and $b$
are chosen to guarantee the conditions $\phi(r_{cut})=0$ and
the first derivative $\phi^\prime(r_{cut})=0$. For the cut-off distance
$r_{cut}=1.2616$ nm the parameter values are $a=0.461$ kJ/mol, $b=-0.313$
kJ/(mol nm) \cite{MNSDST02}.

In the frame of the L-W model of OTP (see Fig. \ref{fig2})
the sites are placed at the vertices of a rigid isosceles triangle. The short
sides of the triangle are of length $\sigma$ and the angle between them is $75^o$.
The position of a site $\alpha$ ($\alpha=1,2,3$) in molecule $i$ is defined by
\begin{equation}
\vec{R}_{i,\alpha}=\vec{R}^{cm}_{i}+\vec{r}_{i,\alpha},
\label{pos}
\end{equation}
where $\vec{R}^{cm}_{i}$ denotes the position of the center of mass of
molecule $i$. In the L-W model OTP molecules are treated as rigid
bodies; each molecule has six (three rotations and three translations) degrees of freedom.
Three of them are taken to be the Cartesian coordinates of the molecule center
of mass (define the vector $\vec{R}^{cm}_{i}$) and other three are the Euler angles
specifying the rotational position about the center of mass (defined by the vectors
$\vec{r}_{i,\alpha}$).

 The energy of interaction between two molecules is given by
\begin{equation}
U_{ij}=\sum\limits_{\alpha=1}^{3}\sum\limits_{\beta=1}^{3}
\phi(\mid \vec{R}_{i,\alpha}-\vec{R}_{j,\beta}\mid).
\label{Utwo}
\end{equation}

\subsection{Simulations details}
We simulated $N=64$ and $N=512$ L-W OTP molecules in a cubic box with
periodical boundary conditions in the frame of the Monte Carlo  method
\cite{MRRTT53}. The trial displacement of a molecule consists of usual random
translation and random rotation (for details of rotations using Euler angles
see \cite{OS77}). The temperature dependence of the particle number density
was estimated by simulations in an isothermal-isobaric (NPT) ensemble
\cite{OS77,W68} with the pressure value fixed at $P=1$ bar. At each
temperature the density was used in a canonical (NVT) ensemble Monte Carlo
simulation. Having results in both ensembles enables us below to compare the specific heats
at constant pressure and at constant volume to experiments.
After short equilibration the average values of the quantities of interest in both ensembles
were measured during $10^7$ sweeps in the case of a system with $N=64$ and
$5\times 10^6$ for $N=512$ molecules. The acceptance rate was chosen to be
$30\%$. A final configuration of MC simulations at constant pressure for
$N=64$ particles in the simulation cell was used to create an initial
configuration for simulation with the larger particle number $N=252$ (by using
eight small cells to initiate one large cell).
Final configurations of MC simulations at constant pressure were used as
initial configurations for simulations at constant volume.
The following dimensionless units are used: the dimensionless length
is  $r^*=r/\sigma$, the particle number density $\rho^*=\rho\sigma^3$,
where $\rho=N/V$ ($V$ is the system volume), the temperature
$T^*=k_B T/\epsilon$, where $k_B$ is the Boltzmann constant, the energy
$U^*=U/\epsilon$, the pressure $P^*=P\sigma^3/\epsilon$ and the specific heat
$C^*=C/(Nk_B)$.  In these dimensionless units $T_m^*=0.58$, $T_b^*=1.07$
and $T_g^*$ is near 0.43 .

\subsection{Simulation results.}

\begin{figure}
\epsfig{width=.38\textwidth,file=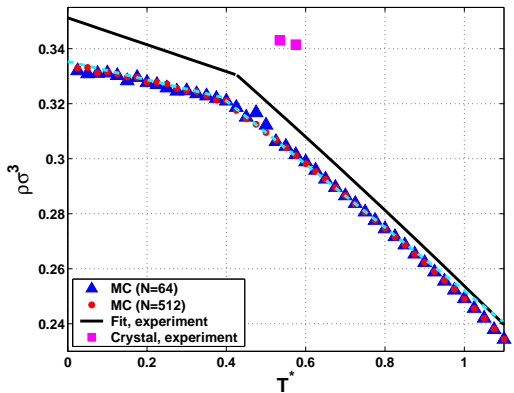}
\caption{Color online: The particle number density as a function of the temperature from
MC simulations at constant pressure. Blue triangles and red dots refer to $N=64$ and $N=512$ respectively. The solid black line shows the fit to the
results of the experimental measurements for OTP (\cite{MFCR01} and references
there). Experimental values for the crystal state of OTP \cite{ODCDES64}
are shown for comparison as pink squares.}
\label{fig3}
\end{figure}

The temperature dependence of the particle number density at constant
pressure is shown in Fig. \ref{fig3}.
Obviously, we can not expect that the simple rigid molecular model of the OTP
molecule should reproduce quantitatively the experimental data. Nevertheless, the
qualitative behavior of the model is similar to the real system; the jump in
the slope associated with the glass transition \cite{GT67} occurs at the
temperature $T_g=0.4$, close to the experimental value. We have not studied
the dependence of  $T_g$ on the thermal protocol of the sample preparation in our simulations
(see, e.g., \cite{PH99}). The results of Fig. \ref{fig3} are independent of
the particle number in the simulation cell. This is an indication of negligible
finite size effects.

\begin{figure}[!h]
\centering
\epsfig{width=.38\textwidth,file=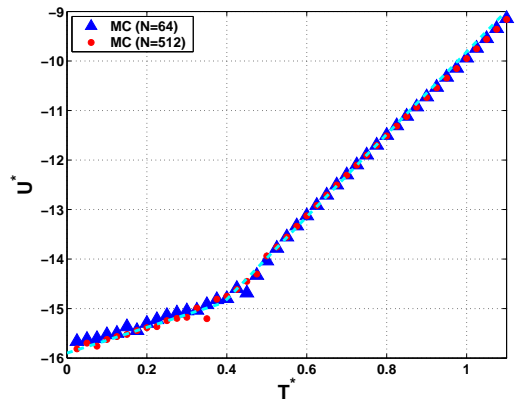}
\caption{Color online: The potential energy of OTP model  as a function of the temperature
from MC simulations at constant volume. }
\label{fig4}
\end{figure}

Fig. \ref{fig4}  shows simulation results of the potential energy change with
the temperature at the same conditions as Fig. \ref{fig3}.
The jump in the slope occurs at the same temperature $T_g=0.4$; again the data are not
sensitive to the particle number in the simulation cell.

\begin{figure}[!h]
\centering
\epsfig{width=.38\textwidth,file=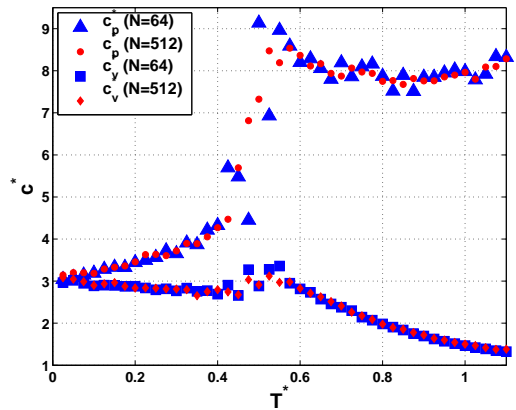}
\caption{Color online: The temperature dependence of the specific heat in the OTP model at
constant pressure and constant volume, such that the volume agrees with
the constant pressure values (see Fig. \ref{fig3}).  }
\label{fig5}
\end{figure}

The specific heat at constant volume, $C_v$, and
constant pressure, $C_p$, are defined by
\begin{eqnarray}
\frac{C_v}{N}&=&\frac{\nu}{2}+\frac{\partial}{\partial T}
\frac{\langle U\rangle}{N}\bigg|_V \nonumber \\
&=&\frac{\nu}{2}+c_v\nonumber\\
\frac{C_p}{N}&=&\frac{\nu}{2}+\frac{\partial}{\partial T}
\frac{\langle U\rangle+PV}{N}\bigg|_P\nonumber \\
&=&\frac{\nu}{2}+c_p,
\label{Cpv}
\end{eqnarray}
where $\nu$ is the number of degrees of freedom.
The 'fluctuation' part of the specific heat is defined at constant volume by
\begin{equation}
c_v=\frac{\langle {U^*}^2\rangle-\langle {U^*}\rangle^2}{N{T^*}^2},
\label{cv}
\end{equation}
where the potential energy $U^*$ is measured in the N-V-T ensemble and
\begin{equation}
c_p=\frac{\langle {H^*}^2\rangle-\langle {H^*}\rangle^2}{N{T^*}^2},
\label{cp}
\end{equation}
where $H^*=U^*-PV$ is measured in the N-P-T ensemble. Simulation results for OTP
are displayed in Fig. \ref{fig5}.

Both specific heats show a drop near the temperature $T_g=0.4$ which is an
indicator of the glass  transition. In experiments one observes qualitative changes in $c_p$ as
a function of the scanning speed \cite{T01}. Our simulations are qualitatively similar
to what is observed in experiments with high scanning speed \cite{GT67}. With
decreasing of the scanning speed the maximum of $c_p$ near the glass
transition point disappears \cite{CB72}. We are not aware of experimental results on
$c_v$ for low temperature OTP; however, similar behavior to ours near the glass
transition point was observed in simulations of binary mixtures
(see \cite{HIP08,HIPS08}).

\begin{figure}[!h]
\centering
\epsfig{width=.38\textwidth,file=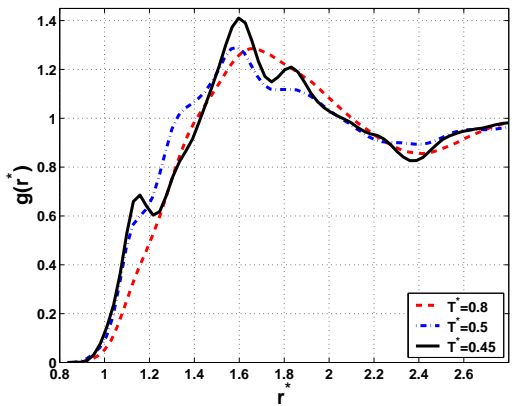}
\caption{Color online: The radial pair distribution function of the OTP center of mass.  }
\label{fig6}
\end{figure}

In order to look for possible structural peculiarities of OTP above $T_g$
we studied the radial pair distribution function of the molecular center of mass
at different temperatures.
Results for three temperatures are shown in Fig. \ref{fig6}. At high
temperatures this function exhibits an ordinary structure typical to liquids
with spherical particles. With decreasing the temperature the position of the first
minimum is unchanged, however, the first peak displays a development of
an internal structure with dips at low temperature. Similar behavior
was detected in Molecular dynamics simulations \cite{LW94}.
This can be attributed to the consequences of nonspherical
intermolecular interactions; at low temperatures and high densities
the system manifests local orientational ordering. Similar evolution
of the pair distribution function was observed in simulations of hard
spherocylinders and it was shown that at high packing fraction there exist short
range orientational \cite{L94}.

Finally we considered the coordination number defined for three dimensional systems as \cite{B77}
\begin{equation}
\langle N\rangle=4\pi\rho\int\limits_{0}^{r_{min}}g(r)r^2\ud r.
\label{Nc}
\end{equation}
Here $r_{min}$ is the position of the first minimum of the pair distribution
\begin{figure}[!h]
\centering
\epsfig{width=.38\textwidth,file=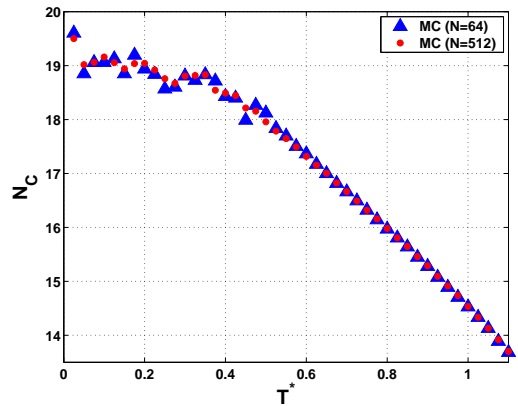}
\caption{Color online: The temperature dependence of the coordination number.  }
\label{fig7}
\end{figure}
function which is one of the possible measure of the range short order. The
temperature dependence of the coordination number is important in
thermodynamic models of the liquid state \cite{SRF80}. The position of
$r_{min}$ is independent of temperature (see Fig. \ref{fig6}), therefore,
another way to estimate the coordination number is to count the average number
of neighbors inside the sphere of the radius $r_{min}$ (numerical result
coincides with the direct integration in Eq. (\ref{Nc})). The result is
displayed in Fig. \ref{fig7}. In the liquid state the temperature dependence
is typical (see \cite{SRF80}). In the vicinity of $T_g$ and below this value
the slope is changed.

\section{Statistical Mechanics of the Glass Transition}
\label{statmech}
\subsection{Statistical geometry approach.}
The structural properties of
condensed matter can be investigated using scattering experiments
(X-rays, electrons or neutrons scattering, see, e.g., \cite{Z79}). The angular
dependence of the scattering intensity is defined by  the structure factor
which is related to the pair distribution function. The scattering experiments
on liquids and amorphous solids exhibit the presence of the short order in
contrast to periodic crystals with long range order.
If in a many-body system energy is presented only by two-body contributions
the estimation of average  thermodynamics values  can be
reduced to  integrals involving  the pair distribution function.
Unfortunately, usually it is impossible to extract the pair distribution
function from experimental data with the desired accuracy. This function is an
average property and in the case of a multi-component systems even the
interpretation of the structure of a system is difficult. Looking for example
at Fig. \ref{fig6} one sees that the radial pair distribution function is not a sufficiently
revealing tool to allow a comprehensive discussion of the glass transition.
As a replacement possible tool we turn now to the analysis of the structure of the
super-cooled liquid in terms of local structures or quasi-species as defined below.

The first suggestion to characterize the state of liquids by local structures
was offered by Bernal
in \cite{B64} in the context of a model of hard balls. The idea is to invoke a geometrical approach to define the liquid
structure. In contrast to a single component crystal with long-range order
where all particles have the same number of neighbors,  a liquid is
random such that the number of neighbors is not fixed, and can vary quite significantly. Defining the concentration
of central particles with $N$ nearest neighbors by $C_N$ with $N\in [N_{\rm min}, N_{\rm max}]$, one understands
that these concentrations will depend on the temperature and pressure,  defining the `state' of the
liquid. This approach, which is referred to as `statistical geometry', does not enjoy an exact
method to calculate the distribution of concentrations $C_N$. On the other hand this distribution is readily
obtained in computer simulations. In a series of papers (see for example \cite{BLPZ09} and references therein) it was shown that for glass formers made of point
particles with soft interactions one can construct an approximate statistical mechanics that allows
a reasonably accurate description of the evolution of the concentrations $C_N$ as a function of the temperature,
with interesting implications for the study of the glass transition. In this paper we extend this approach
to molecular liquids as explained below. The advantage of this approach (see for example \cite{F64}) compared to say, approaches based on the pair distribution function, is that once the statistical mechanics is set up one can compute thermodynamic
properties in the usual way that is found in any textbook on statistical physics.

\subsection{Quasi-species in Supercooled Molecular Liquids}
The crucial step in constructing an approximate theory is the identification of appropriate quasi-species. In the case of
molecular liquids this task is not immediately obvious. As said above, we want to identify
quasi-species which are made of a central molecule and a number of nearest neighbors, and these should have well defined configurations with well defined energies. The crucial assumption is that a given quasi-species interacts weakly with other quasi-species compared to the interaction of the central molecule with its nearest neighbors. As attractive as this this idea
may sound, there are practical problems which appear if the number of such quasi-species
required to describe the liquid configurations is not small. For example a non spherically symmetric molecule can take up many orientations relative to a given central molecule. Each orientation will result in a different energy of interaction. Take for example a situation in which the central molecule can have up to $N_{max}$ neighbors, and each neighbor molecule can be in two orientations. Then we already have $2^{N_{max}}$ possible configurations, each in principle having a different energy. Thus if $N_{max}=35$ we might have $\approx 3\times 10^{10}$ quasi-species--a ludicrous number that destroys the entire concept of quasi-species. To simplify things towards a workable theory (that will have to be validated against numerical data) we will treat all nearest neighbor sites as being identical, and the molecules occupying these sites as having $m$ internal degrees of freedom (for example orientations relative to the central molecule) with energies $\epsilon_1,\dots,\epsilon_m$. The total energy of the quasi-species will be estimated as the
sum of the energies of interaction of the neighbors with the central molecule. The interaction between the neighbors themselves
will be estimated below as a 'crowding effect' that will be characterized by a single additional parameter, following ideas
prevalent in polymer physics. We turn now to a making these ideas concrete.

\subsection{Statistical Mechanics of Quasi-species Concentrations}

We begin by assuming that the neighboring molecules are not interacting with each other. In this case
every neighboring molecule can have an energy taken from $m$ possible discrete values $\epsilon_1,\cdots,\epsilon_m$.
Denote by $N_1$ the number of ways that we can fit molecules with energy $\epsilon_1$, consistent with $N_2$ ways to fit
molecules with energy $\epsilon_2$ etc. such that in total $N_1+N_2+\cdots+N_m=N$ where $N\le N_{max}$ and $N_{max}$ is the maximum number of possible nearest neighbors. Then the number of such configurations is
\begin{eqnarray}
\label{a}
\Omega(N_1,\dots,N_m) &=& \frac{N_{max}!}{(N_{max}-N)! N!}
\frac{N!}{\Pi_{k=1}^m N_k!} \nonumber \\
& = &  \frac{N_{max}!}{(N_{max}-N)! \Pi_{k=1}^m N_k!} .
\end{eqnarray}
As the energy of such a configuration is
\begin{equation}
\label{b}
U(N_1,\dots,N_m) = \sum_{k=1}^m N_k \epsilon_k ,
\end{equation}
the Boltzmann weight of such a configuration  is
\begin{equation}
W(N_1,\dots,N_m) =\Omega(N_1,\dots,N_m) e^{-\frac{\displaystyle{1}}
{\displaystyle{k_BT}} \displaystyle{\sum_{k=1}^m N_k
\epsilon_k}}
\label{b1}
\end{equation}
and consequently the total weight of any $N$ nearest neighbors
is
\begin{equation}
\label{c}
W(N) = \sum_{N_1}\dots \sum_{N_m} W(N_1,\dots,N_m)
\delta_{N,\sum_{k=1}^m N_k},
\end{equation}
where $\delta_{i,j}$ is the Kronecker delta.
Using the identity
\begin{eqnarray}
\label{d}
&&\bigg[\sum_{k=1}^m e^{-\frac{\displaystyle{1}}{\displaystyle{k_BT}}
\displaystyle{\epsilon_k}}\bigg]^N = \sum_{N_1}\dots \sum_{N_m}
\frac{N!}{\Pi_{k=1}^m N_k!} \nonumber \\
&\times&e^{-\frac{\displaystyle{1}}{\displaystyle{k_BT}}
\displaystyle{\sum_{k=1}^m N_k \epsilon_k}}
\delta_{N,\sum_{l=1}^m N_l}
\end{eqnarray}
Eq. (\ref{c}) can be written explicitly as
\begin{eqnarray}
\label{c1}
W(N) & = & \sum_{N_1}\dots \sum_{N_m}  \frac{N_{max}!}
{(N_{max}-N)! \Pi_{k=1}^m N_k!} \nonumber \\
&\times&e^{-\frac{\displaystyle{1}}{\displaystyle{k_BT}}
\displaystyle{\sum_{k=1}^m N_k \epsilon_k}}
\delta_{N,\sum_{l=1}^m N_l}\nonumber \\
& = &  \frac{N_{max}!}{(N_{max}-N)! N!}
\bigg[\sum_{k=1}^m e^{-\frac{\displaystyle{1}}{\displaystyle{k_BT}}
\displaystyle{\epsilon_k}}\bigg]^N.
\end{eqnarray}

Next we need to consider the interaction between neighboring molecules.
Clearly, the assumption that the energy of a configuration is defined by
Eq. (\ref{b}) is correct only for $N\ll N_{max}$. The most obvious physical
effect that destroys this assumption is the soft-core repulsive interaction that will exist between
nearest neighbor molecule, especially as $N$ approaches $N_{max}$. To
include such repulsive forces
exactly is extremely cumbersome. In principle, if all nearest
neighbor sites are not identical, the repulsive energies can vary
between different pairs of nearest neighbor molecules. Even if these
energies are identical the total repulsive energy will depend on the exact
placement of $N$ nearest neighbor molecules in the available $N_{max}$ sites.
Such considerations go beyond our desire to simplify the model as much as possible.
We propose therefore a mean-field
approximation in which the repulsive energy is described by only one new
parameter.
The reasoning is similar to that used in the polymer physics \cite{Z79,G79}.
Suppose that there exists a soft core repulsive energy $J$ between {\it each pair}
of nearest neighbor molecules in the first shell surrounding a central
molecule.  The probability that a nearest neighbor site is occupied is $N/N_{max}$
and therefore there will exist an additional term to the energy given by Eq.
(\ref{b}) due to this repulsive contribution,
\begin{eqnarray}
\label{g}
U_{rep}&=& \frac{1}{2}J \frac{N}{N_{max}} N \nonumber \\
&=& J \frac{N^2}{2 N_{max}},
\end{eqnarray}
which is quadratic in $N$.
Therefore, Eq. (\ref{c1}) is modified by the multiplication with the term
$exp(-\frac{J}{k_BT}\frac{N^2}{2 N_{max}}) $ and the concentration of quasi
particles with $N$ nearest neighbours reads
\begin{equation}
C_N=\frac{N_{max}!}{(N_{max}-N)! N!}
\frac{\exp(\ln S(T)N-\frac{J}{k_BT}\frac{N^2}{2 N_{max}})}{Z(T)},
\label{Cn}
\end{equation}
where the partition function is
\begin{eqnarray}
\label{i}
Z(T) &=& \sum_{N=1}^{N_{max}} \frac{N_{max}!}{(N_{max}-N)! N!}\nonumber \\
&\times&\exp\bigg(\ln S(T)N-\frac{J}{k_BT}\frac{N^2}{2 N_{max}}\bigg)
\end{eqnarray}
and
\begin{equation}
S(T)=\sum_{k=1}^m e^{-\frac{\displaystyle{\epsilon_k}}{\displaystyle{k_BT}} }.
\label{S}
\end{equation}

\subsection{Numerical results.}
The most precise way to define `nearest neighbors' is via the
construction of a Voronoi tessellation (of polyhedra or polygons in three-
or two- dimensional space correspondingly) \cite{V908}. In the frame of the
Bernal approach this method was used in MC simulations to study the statistical
geometry of the Lennard-Jones liquid \cite{FI70,FII70}.
\begin{figure}[!h]
\centering
\epsfig{width=.38\textwidth,file=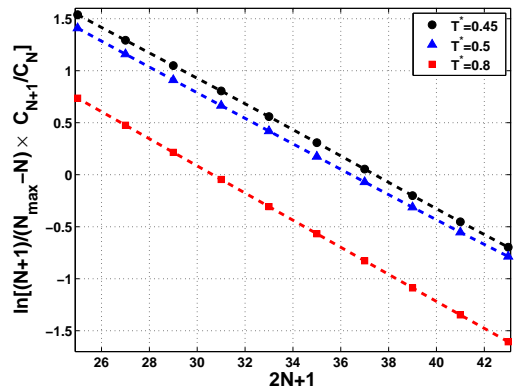}
\caption{Color online: Chekup of Eq.~\ref{f} ($N_{max}=35$).}
\label{fig8}
\end{figure}
Unfortunately, the computational cost of a Voronoi construction is very
high. A simpler definition of `nearest neighbors' in MC or MD simulations
was proposed in \cite{SNR83}. The nearest neighbors of a given particle are
considered as particles with their centers inside the first coordination
shell.
\begin{figure}[!h]
\centering
\epsfig{width=.38\textwidth,file=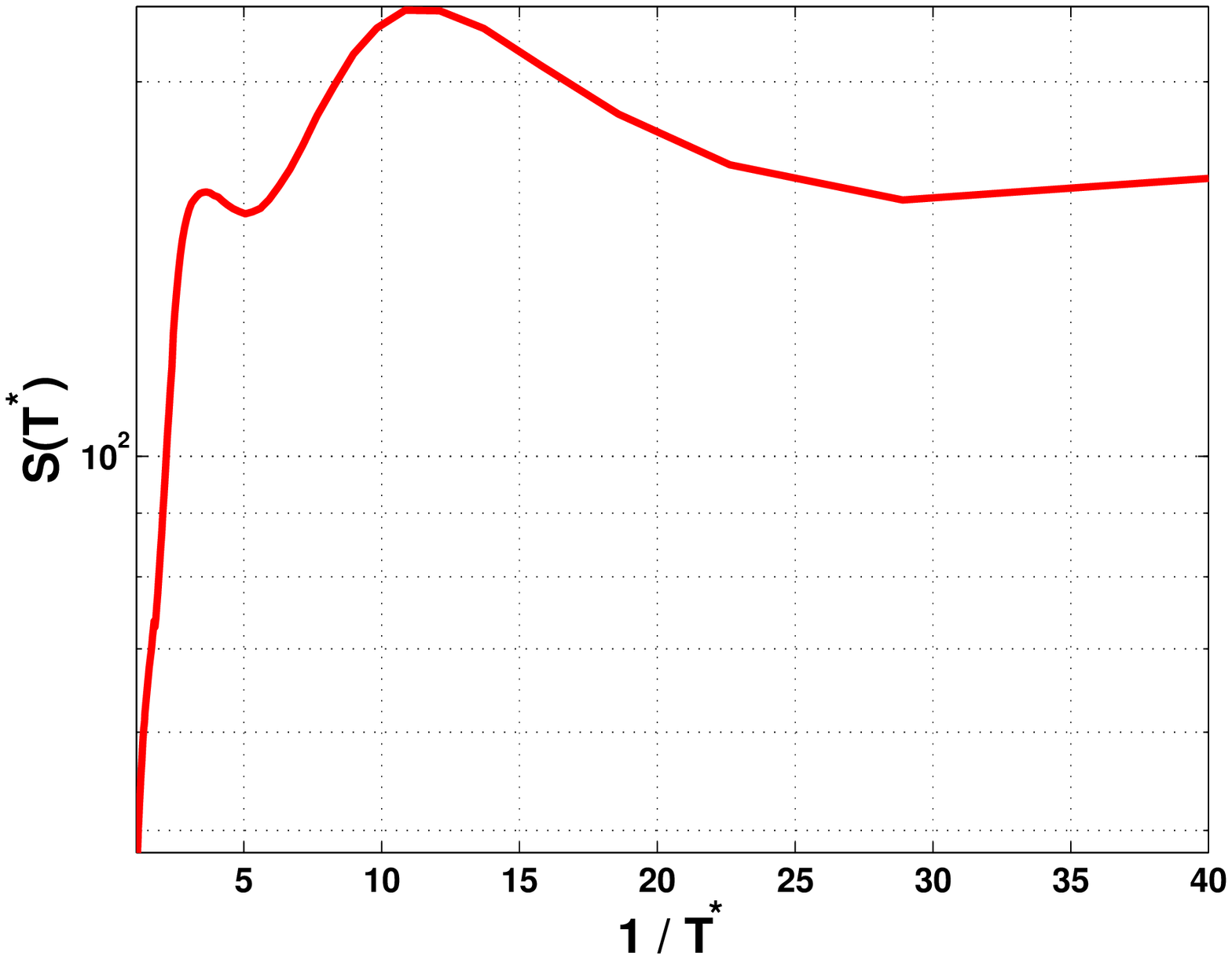}
\epsfig{width=.38\textwidth,file=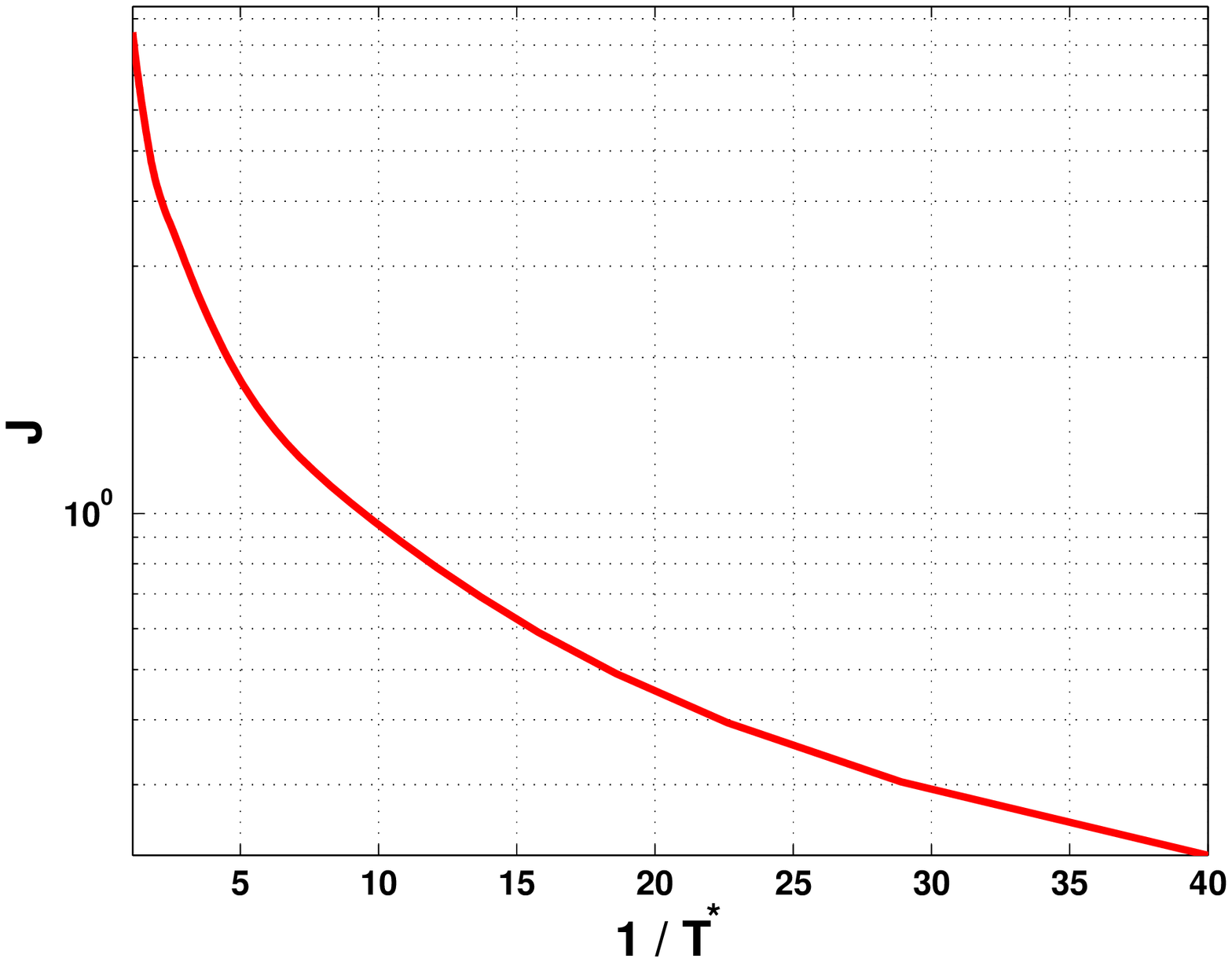}
\caption{Color online: Upper panel: The function $S(T^*)$ defined as the intercept
in lines from Fig. \ref{fig8}. Lower panel: The parameter $J$defined as the
slope in lines from Fig. \ref{fig8}  .}
\label{fig9}
\end{figure}
It  was shown that if the second coordination shell is excluded the
results are not sensitive to details of the  nearest neighbors definition.
Therefore, we counted a particle as a nearest neighbor of a central one if
the distance between their centers of mass is less then $r_{min}$ in
Eq.~\ref{Nc}. The result of this calculation is given in terms of the temperature dependence of the
concentration distribution $C_N$.
\begin{equation}
\label{f}
\ln{[\frac{(N+1)C_{N+1}}{(N_{max}-N)C_{N}}]} =
\ln S(T)-\frac{J}{k_BT}\frac{2N+1}{2 N_{max}} \ .
\end{equation}
This suggests that if $\ln{[\frac{(N+1)C_{N+1}}{(N_{max}-N)C_{N}}]}$
is plotted against $2N+1$
there should result a straight line of slope $-\frac{J}{k_BT}\frac{1}{2 N_{max}}$
and intercept $\ln S(T)$. Typical simulations results and comparison with the prediction of Eq.~\ref{f}
are shown in Fig. \ref{fig8}. The slope and the intercept calculated
by a least squares fit for different temperatures are plotted in
Fig. \ref{fig9}.

From Eq. (\ref{S}) the function $S(T)$ is given as a sum of exponentials. It turns out that it is very difficult to fit $S(T)$ as a sum
of a small number of exponentials with temperature independent energies $\epsilon_k$. This may result from the fact that there are
many such energies, or, as these are in fact phenomenological parameters, they are not at all
guaranteed to be temperature independent, since they encapsulate information about the complicated interactions in the system. The same is true for the parameter $J$. Luckily, in the frame of the simple model
proposed here we do not need the individual parameters $\epsilon_k$ since they always appear in the same combination in the form of the single function $S(T)$ (see Eq.~\ref{Cn}). This function together with  $J(T)$, both of which can be estimated from the
simulations, are all that we need to determine the temperature dependence of the concentrations $C_N(T)$ which are the crucial
observables in the present approach.

\begin{figure}[!h]
\centering
\epsfig{width=.38\textwidth,file=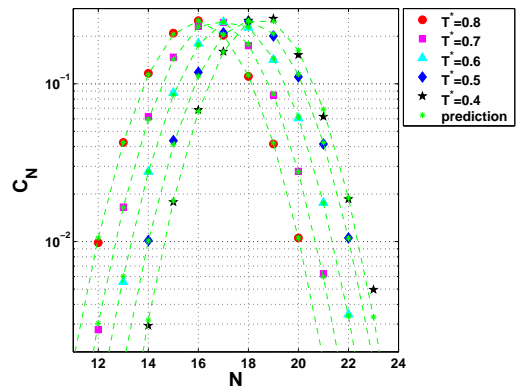}
\caption{Color online: Temperature and $N$ dependence of $C_N(T)$. Symbols are the direct numerical simulations and dashed lines are correspond
to the model prediction, Eq.~\ref{Cn} .}
\label{fig10}
\end{figure}

\begin{figure}[!h]
\centering
\epsfig{width=.38\textwidth,file=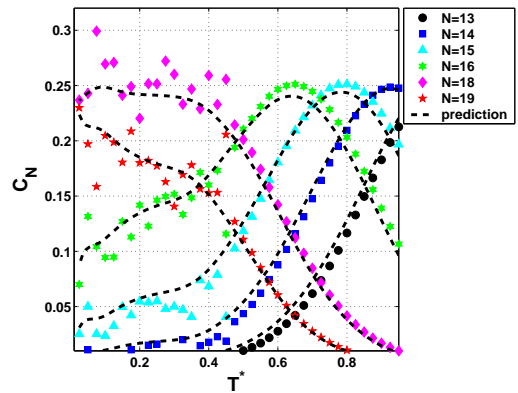}
\caption{Color online: Dependence of the quasi-species concentration  on the temperature
at fix number of
near neighbor particles. The dashed lines correspond to the model prediction Eq.~\ref{Cn} .}
\label{fig11}
\end{figure}

Having determined $S(T)$ and $J(T)$ from the numerics (cf. Fig. \ref{fig9}), the model prediction of the temperature dependence of $C_N(T)$ is shown in Fig, \ref{fig10}, in comparison with the direct numerical simulation results. The agreement appears excellent.
It is interesting to comment here that similar Gaussian looking distributions for a given temperature as a function of $N$
was found in all the models of glass formers that were studied using a similar quasi-species approach \cite{BLPZ09}. Close to the dependence displayed in
Fig. \ref{fig10} was observed in simulations of water model \cite{MSA11}.  We believe that this
is a generic feature that will be common to a very wide class of liquids and glass formers once the appropriate quasi-species
are identified. The difference between one system and the other will be encoded in the temperature dependence of these
Gaussian shapes. To see this we turn in Fig. \ref{fig11} to a comparison of the direct numerical simulations to the model predictions for the temperature dependence of $C_N(T)$ at chosen values of $N$.
This may be the most relevant encoding of the subtle changes in the structure of the liquid as temperature is changed.
We see (quite generically again) that some quasi-species decay in their concentration when temperature is decreased,
some increase in concentration, and some first increase and then decrease to a finite value as $T\to 0$.
The quasi-species whose concentration tends to zero are referred to here and in previous work as `liquid-like', since
their concentration is significant only in the high-temperature liquid. These are those associated with the highest free
energy (made from both energy and degeneracy) and therefore their concentration declines when the temperature decreases.
We will argue that they carry with them many of the signatures of the glass transition.

Like in previous studies \cite{ILLP07} it is worthwhile to group together all the quasi-species into three distinct groups
according to their qualitative temperature dependence (decreasing, increasing, and first increasing and then decreasing).
We refer to the resulting model as a `Doubly Coarse Grained' (DCG) description.
In the present case this means summing the following three groups
\begin{equation}
C_I\equiv \sum_{N=2}^{14} C_N(T)\ , C_{II}\equiv \sum_{N=15}^{17} C_N(T)\ ,  C_{III}\equiv\sum_{N=18}^{35} C_N(T) \ .
\label{groups}
\end{equation}
\begin{figure}[!h]
\centering
\epsfig{width=.38\textwidth,file=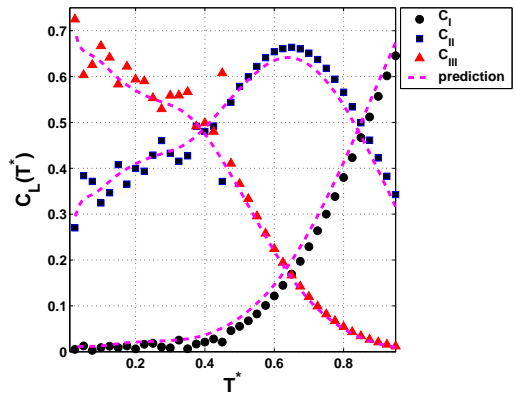}
\caption{Color online: Comparison of MC simulations with the model prediction for
grouped concentrations (see Eq.~\ref{groups}).}
\label{fig12}
\end{figure}
The temperature dependence of these three groups are shown in Fig. \ref{fig12}, again comparing the model predictions to the direct numerical simulation with
an obvious excellent agreement. One can see that the glass transition
temperature $T_g\sim 0.4$ corresponds to the intersection point of
$C_{II}(T^*)$ and $C_{III}(T^*)$, at this point $C_I(T^*)$ practically dies
out.

\begin{figure}[!h]
\centering
\epsfig{width=.38\textwidth,file=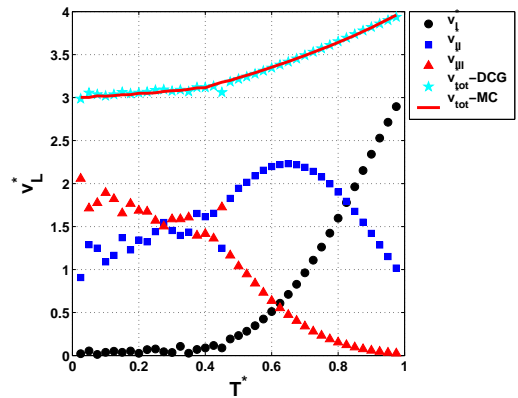}
\caption{Color online: Volume contributions of the quasi-species.}
\label{fig14}
\end{figure}

The price of the `nearest neighbors' definition used by
us in contrast to the construction of a Voronoi tessellation is the loss
of the information on a volume ascribed to a quasi-species. Nevertheless, one
can try to use the definition of the perfect solution molar volume \cite{P57}
\begin{eqnarray}
v^*_{tot}&=&\sum\limits_{L=I}^{III} v^*_{L}(T^*)
\nonumber \\
&=&\sum\limits_{L=I}^{III}\langle v^*\rangle_{L}C_{L}(T^*),
\label{mVol}
\end{eqnarray}
where $v^*_{tot}=1/\rho^*$.
The partial volumes $\langle v^*\rangle_{L}$ were estimated by least squares
fits in the whole temperature range and are displayed in Table~\ref{tab1}.
\begin{table}[!h]
\centering
\caption{Color online: Partial volumes $\langle v^*\rangle_L$,
densities $\rho^*_L$ and
energies $\langle U^*\rangle_L$.}
\begin{tabular}{|c|c|c|c|}
\hline
L&$\langle v^*\rangle_L$&$\rho^*_L$&$\langle U^*\rangle_L$ \\
\hline
I&4.204&0.238&-9.13 \\
II&3.651&0.298&-12.49 \\
III&2.839&0.352&-17.29 \\
\hline
\end{tabular}
\label{tab1}
\end{table}
The corresponding densities $\rho^*_L=1/\langle v^*\rangle_L$ are also presented in
this table;
one can see from the comparison with the data of Fig. \ref{fig3} that
the  value of $\rho^*_{III}$ is close to the density of the crystal.
Results for  the temperature
dependence of $v^*_{L}(T^*)$ are shown in Fig. \ref{fig14}. The
estimation of the molar volume by Eq.~(\ref{mVol}) is compared with MC data.

\begin{figure}[!h]
\centering
\epsfig{width=.38\textwidth,file=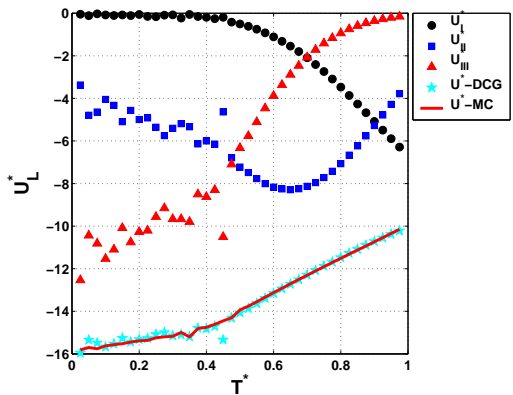}
\caption{Color online: Energy contributions of the quasi-species.}
\label{fig15}
\end{figure}

In the frame of the approximation of the perfect solution the potential
energy is given by
\begin{eqnarray}
U^*&=&\sum\limits_{L=I}^{III} U^*_{L}(T^*)
\nonumber \\
&=&\sum\limits_{L=I}^{III}\langle U^*\rangle_{L}C_{L}(T^*),
\label{mEn}
\end{eqnarray}
The coefficients $\langle U^*\rangle_{L}$ are estimated again by least squares
fits in the whole temperature range and are displayed in Tab.~\ref{tab1}.
The contributions to the potential energy and the comparison of the potential
energy from simulations and the prediction of Eq. (\ref{mEn}) are shown in
Fig. \ref{fig15}).
The  agreement of MC results and predictions of Eq. (\ref{mVol}) and
Eq. (\ref{mEn})  means that in spite of the strong
assumption of additivity in these equations the mixture of quasi-species
can be effectively considered as a perfect solution.

\begin{figure}[!h]
\centering
\epsfig{width=.38\textwidth,file=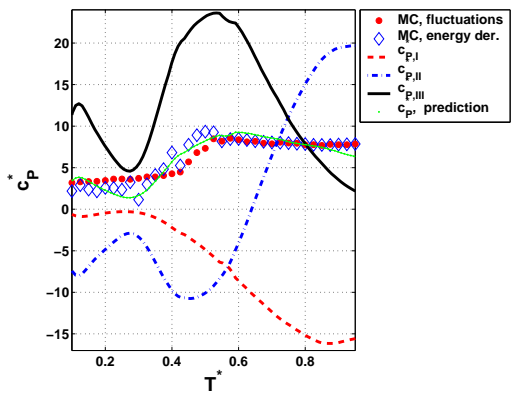}
\caption{Color online: Contributions of the quasi-species to the heat capacity.}
\label{fig16}
\end{figure}

The enthalpy can be connected with the quasi-species concentration using
Eq. (\ref{mVol}) and Eq. (\ref{mEn}). Numerical differentiation of
concentrations given by Eq. (\ref{groups}) yields the quasi-species
contributions (see Eq. (\ref{Cpv})). Results of these calculations and
comparison with numerical differentiation of the enthalpy estimated in MC
runs and data from Fig. \ref{fig5} is displayed in Fig. \ref{fig16}.
It follows from Fig. \ref{fig14}-Fig. \ref{fig16} that the glass transition
temperature defined by volumetric and thermal measurements can be
associated with the disappearance of the quasi-species of kind $I$.

\section{Local structures and viscosity of OTP}
\label{viscosity}
The theory of the viscosity of gases and liquids saw a long and confusing
history, especially for super-cooled liquids, \cite{HCB54,B13,M23,H71,EZ72,A30,V21,F25,T25,TH26,MBCLSCS10,E36,AG65,RA98,G11,A05,HNOD08,SM09,A11}. For our
purposes we refer to the interpolation formula that was proposed in \cite{D51,D57,CT59}
\begin{equation}
\eta=\eta_{\infty}e^{A/v_f}.
\label{dool}
\end{equation}
Here $v_f$ is the free volume that in general is not easy to assign to any physical object.
The idea behind this interpolation formula is that for particles to move out of their immediate cages some
free volume needs to open up to allow flow. In our context it can be assumed that
the decline in concentration of $C_{I}(T^*)$ is responsible for the increase in viscosity;
the free volume in this case is defined as $v^*_f=v^*_I(T^*)$ (see also
\cite{ABHIMPS07,HIMPS07}). In order to confirm this suggestion it is
necessary to consider the temperature dependence of the model free volume
and the measured viscosity from the point of view of Eq. (\ref{dool}).

The temperature dependence of the dimensionless volume $v^*_I(T)$ shown in
the upper panel of Fig. \ref{fig18} above the glass transition temperature
$T_g$ can be approximated by
\begin{equation}
\ln v^*_I(T)=3.3079-\frac{5.95T_g}{T}.
\label{vappr}
\end{equation}

\begin{figure}[!h]
\centering
\epsfig{width=.38\textwidth,file=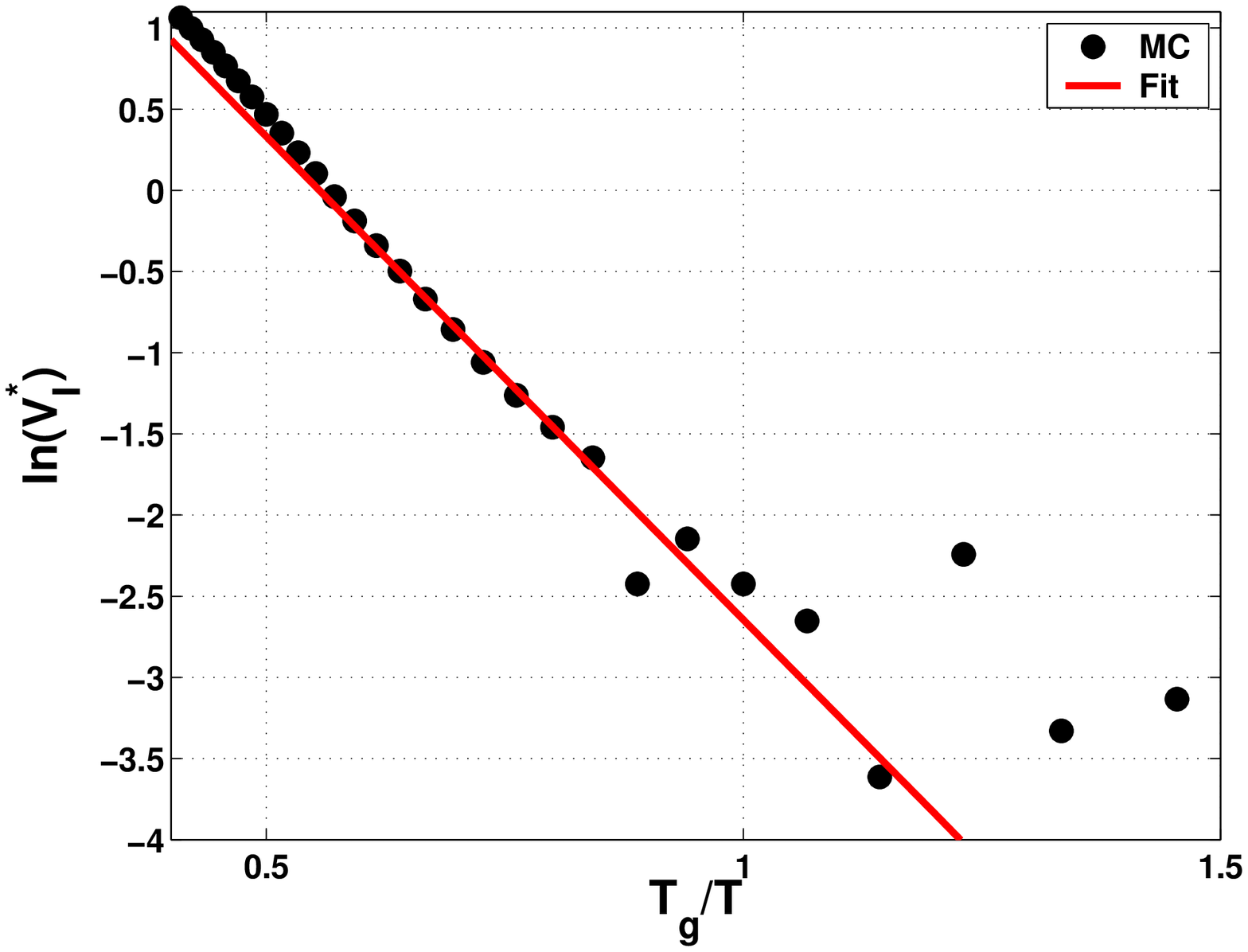}
\epsfig{width=.38\textwidth,file=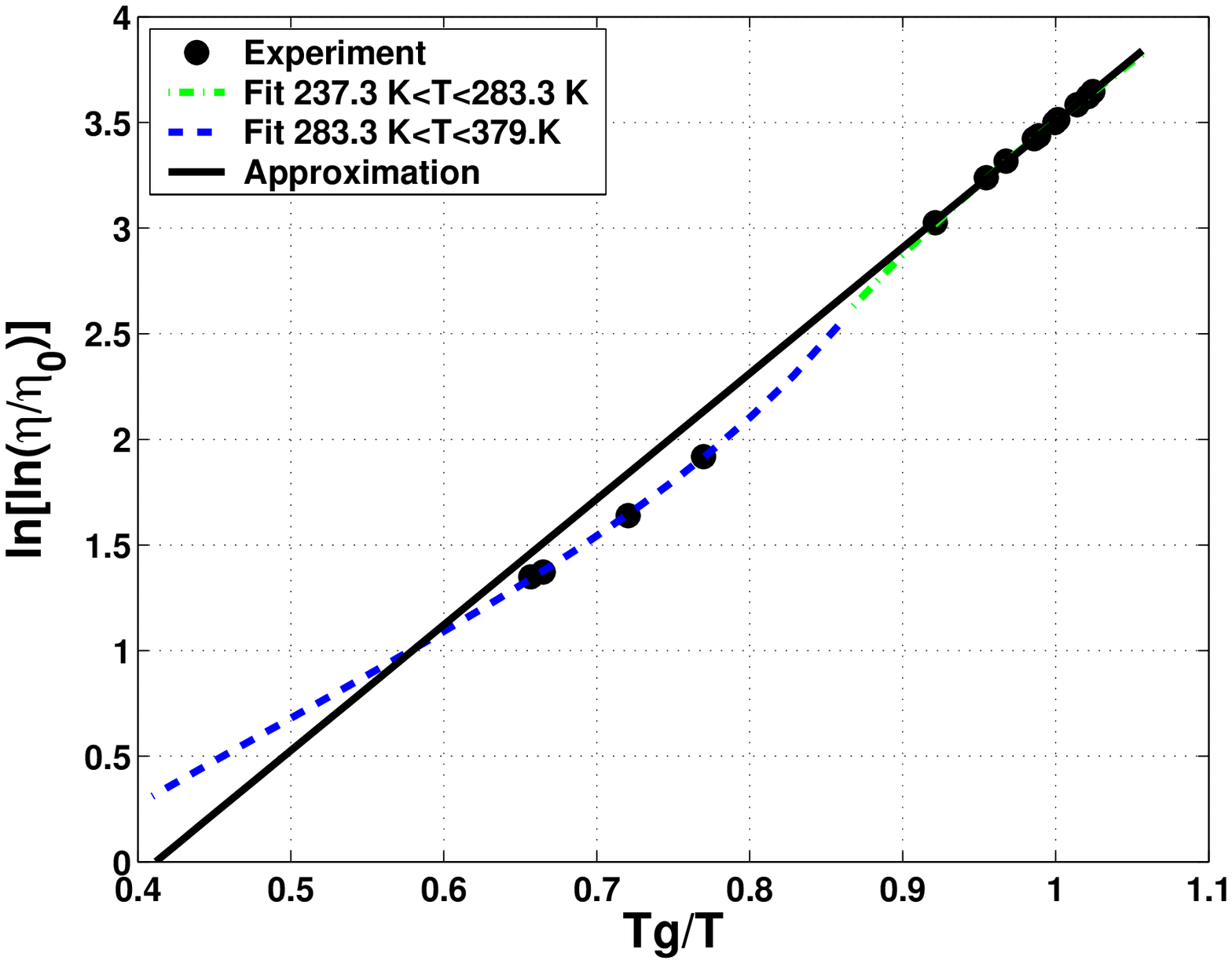}
\caption{Color online: Upper panel: Temperature dependence of the volume $v^*_I(T)$.
Lower panel:Temperature dependence of the experimental OTP viscosity \cite{PCC94}
(the value of $\eta_\infty$ is taken from \cite{CE93}).}
\label{fig17}
\end{figure}

The linearized temperature dependence of the measured OTP viscosity
\cite{PCC94} is
displayed in the lower panel of Fig. \ref{fig17}. In the vicinity of the glass
transition temperature it is approximated by
\begin{equation}
\ln \ln \frac{\eta}{\eta_\infty}=-2.45+\frac{5.95T_g}{T},
\label{etappr}
\end{equation}
where $\eta_{\infty}=1.25\times 10^{-3}$ poise is taken from
\cite{CE93} for $T > 283.3 K$.
Combination of Eq. (\ref{vappr}) and Eq. (\ref{etappr}) yields the free
volume expression in the following form
\begin{equation}
\eta=\eta_{\infty}\exp{(v^*_c/v^*_I(T)})\ ,
\label{cfrv}
\end{equation}
where $v^*_c=2.36$.

In Fig, \ref{fig18} we present the predictions of the last equation
(\ref{cfrv}) to the experimental measurements of the viscosity in OTP as a
function
of $T/T_g$. The noise in our measurement of $C_I(T)$ and $v_I(T)$
(cf Figs. 12,13) is
reflected in the scatter of the predicted viscosity. If we use the model
prediction for
$C_I(T)$ and $v_I(T)$ we find the solid red line.
\begin{figure}[!h]
\centering
\epsfig{width=.38\textwidth,file=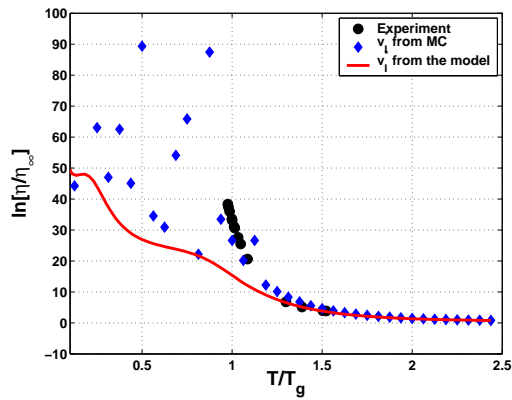}
\caption{Color online: Comparison of the experimental measurements of the OTP viscosity
\cite{PCC94} with the predictions of Eq. (\ref{cfrv}). Blue rhombi correspond
to MC values of concentrations in Eq. (\ref{groups}), the solid red line
represents calculation results with concentrations predicted by the
model given by Eq. (\ref{Cn}).}
\label{fig18}
\end{figure}
It follows from this figure that proposed definition of the free volume yields
good agreement with the experimental data in the whole temperature range using
the concentration as measured. The model prediction underestimates
the viscosity at low temperatures;   this must follow from a discrepancy between
the model prediction and the actual simulational concentrations at these temperatures. Nevertheless, the simulation data
indicates that below the glass transition temperature the rate of increase
of the viscosity is slowing down. This conclusion agrees with the data
in \cite{KTH00}.
\section{Discussion.}
\label{discussion}

The basic assertion of this paper is that super-cooled liquids, be them simple or molecular liquids, are ergodic
systems that are describable by statistical mechanics. It is quite impossible to write and to solve for the full
partition function that takes into account all the interactions and the degrees of freedom in these systems.
We have therefore developed a program in which we offer an approximate statistical mechanics that is based on
considering, for every particle in the system, only the interactions with the nearest neighbors. This defines
our `quasi-species' which consist of a central particle and its $N$ neighbors, where $N$ can vary quite a bit.
In many cases it turns out also advantageous to group the quasi-species into groups, such that one
group consists of all the quasi-species whose concentrations decline when $T$ reduces. Usually another group contains
all the quasi-species whose concentration increase as $T$ decreases, with a third group whose concentrations neither
increases nor decreases \cite{BLPZ09}. We referred to the resulting model as the DCG model, in which the concentration
of the first group acts as an order-parameter that signals the glass transition. In other words, since the quasi-species that
disappear when $T\to 0$ are those of highest free energy, we consider them as them as liquid-like. In previous papers
\cite{BLPZ09} we considered the inverse concentration of this group as a typical scale measuring the distance between
the quasi-species raised to the appropriate power. Here we opted to compute numerically the volume associated with these
disappearing objects and considered this as the `free volume' of Eq. (\ref{dool}). Using this we could estimate the
viscosity, and the results were compared to experiments in Fig. \ref{fig18}. It is quite gratifying to see the agreement
between the predicted viscosity (blue rhombi) and the experimental data in the regime below the glass transition.
We do not have data for $T<T_g$ from the experiment, but we still have (admittedly quite scattered) data for the functions $S(T)$ and $J(T)$ from which
we can estimate the concentrations of the quasi-species. Using these we predict that the viscosity remains finite, even though extremely large (note the logarithmic scale in Fig. \ref{fig18} ), even for $T\to 0$. We stress however that what really happens
at $T=0$ cannot be safely deducted from the results of this study; it is much better to consider the theory of elasticity
at $T=0$ as can be found in Ref. \cite{11HKLP}.

\end{document}